\documentclass[a4paper,notitlepage,12pt]{article}
\usepackage{amsfonts}
\usepackage{amsmath}

\input{tcilatex}

\begin{document}

\title{Bender--Dunne Orthogonal Polynomials, Quasi--Exact Solvability and
Asymptotic Iteration Method for Rabi Hamiltonian}
\author{\textbf{S.--A. Yahiaoui}\thanks{%
E-mail: s\_yahiaoui@univ-blida.dz}, \textbf{M. Bentaiba}\thanks{%
E-mail: bentaiba@univ-blida.dz} \\
LPTHIRM, D\'{e}partement de Physique, Facult\'{e} des Sciences,\\
Universit\'{e} Saad DAHLAB de Blida, Blida, Algeria}
\maketitle

\begin{abstract}
We present a method for obtaining the quasi--exact solutions of the Rabi
Hamiltonian in the framework of the asymptotic iteration method. The energy
eigenvalues, the eigenfunctions and the associated Bender--Dunne orthogonal
polynomials are deduced. The latter prove to have a nonpositive definite
norm.


\end{abstract}

\section{Introduction: formal aspects}

A new class of potentials which are intermediate to exactly solvable (ES)
ones and non solvable ones are called quasi--exactly solvable (QES) problems 
\cite{1,2,3,4}. It is well known that a part of their spectrum can be
determined algebraically but not whole spectrum. One of an alternative
mathematical language for describing the QES systems is a remarkable set of
orthogonal polynomials introduced by Bender and Dunne \cite{5}. These
polynomials satisfy the three--term recursion relation and, as a
consequence, form an orthogonal set with respect to some weight function $%
\rho \left( \mathcal{E}\right) $ \cite{6,7}. The exact solution of the Schr%
\"{o}dinger equation takes the form:%
\begin{equation}
\chi \left( x;\mathcal{E}\right) =\sum\limits_{n=0}^{\infty }\chi _{n}\left(
x\right) \mathcal{P}_{n}\left( \mathcal{E}\right) ,  \tag{1}
\end{equation}%
in which $\mathcal{P}_{n}\left( \mathcal{E}\right) $ denotes certain
polynomials of $\mathcal{E}$ and satisfy \cite{8,9}:%
\begin{equation}
\mathcal{P}_{n}\left( \mathcal{E}_{k}\right) =0;\quad n\geq \Lambda +1,\quad
k=0,1,2,\cdots ,\Lambda .  \tag{2}
\end{equation}%
where the parameter $\Lambda $ is a positive integer value. It follows that $%
\mathcal{P}_{n}\left( \mathcal{E}\right) $ are orthogonal:%
\begin{equation}
\int \mathcal{P}_{n}\left( \mathcal{E}\right) \rho \left( \mathcal{E}\right) 
\mathcal{P}_{m}\left( \mathcal{E}\right) d\mathcal{E}=p_{n}p_{m}\delta _{nm},
\tag{3}
\end{equation}%
where $p_{n}$ represent the norms of the orthogonal polynomials $\mathcal{P}%
_{n}\left( \mathcal{E}\right) $. It is possible to determine the norms $p_{n}
$ (or squared norms $\gamma _{n}=p_{n}^{2}$) by multiplying the recursion
relation by $\mathcal{E}^{n-1}\rho \left( \mathcal{E}\right) $ and to
integrate with respect to $\mathcal{E}$.

However, the search of the orthogonal polynomials $\mathcal{P}_{n}\left( 
\mathcal{E}\right) $ can be generated by means of the asymptotic iteration
method (AIM) \cite{10}. This procedure was first introduced as an
approximation in order to deduce both energy eigenvalues and eigenfunctions,
using some computer--algebra systems. Due to the simplicity of the
procedure, various aspects of the method have been employed, successfully,
to obtain both ES and QES problems \cite{10,11,12,13,14}. The systematic
procedure of the AIM begins by rewriting a second--order linear differential
equation in the following form:%
\begin{equation}
\chi ^{\prime \prime }\left( x\right) =r_{0}\left( x\right) \chi ^{\prime
}\left( x\right) +s_{0}\left( x\right) \chi \left( x\right) ,  \tag{4}
\end{equation}%
with $r_{0}\left( x\right) $ and $s_{0}\left( x\right) $ are many times
differentiable functions. In order to find the solution of Eq.(4), we rely
on the symmetric shape of the right--hand side. Thus for $\left( n+1\right)
^{\text{th}}$ and $\left( n+2\right) ^{\text{th}}$ derivative of Eq.(4), $%
n=1,2,\cdots $, we get:%
\begin{eqnarray}
r_{n}\left( x\right) &=&r_{n-1}^{\prime }\left( x\right) +s_{n-1}\left(
x\right) +r_{0}\left( x\right) r_{n-1}\left( x\right) ,  \TCItag{5.1} \\
s_{n}\left( x\right) &=&s_{n-1}^{\prime }\left( x\right) +s_{0}\left(
x\right) r_{n-1}\left( x\right) ,  \TCItag{5.2}
\end{eqnarray}%
where for sufficiently large iteration number $n$, the \textquotedblleft 
\textit{asymptotic}\textquotedblright\ behavior of the procedure can be
applied as:%
\begin{equation}
\frac{r_{n-1}\left( x,\mathcal{E}\right) }{s_{n-1}\left( x,\mathcal{E}%
\right) }=\frac{r_{n}\left( x,\mathcal{E}\right) }{s_{n}\left( x,\mathcal{E}%
\right) }=\alpha \left( x,\mathcal{E}\right) ,  \tag{6}
\end{equation}%
and which allows us, on the one hand, to calculate the energy eigenvalues by
iterations and, on the other hand, to express the so--called polynomial
solutions of Eq.(4).

Many interesting models are obtained by combining two fundamental models of
quantum mechanics; namely the two--level system and the harmonic oscillator 
\cite{15}. The Rabi Hamiltonian is one of the more studied ones \cite{16,17}%
, and plays an important role in many areas of physics from condensed matter
and biophysics to quantum optics. Due to the elegance of the model various
aspects of the Hamiltonian have been studied analytically and numerically 
\cite{18,19,20,21,22,23}. The complete Hamiltonian of a such system is of
the form:%
\begin{equation}
H=\hbar \omega a^{\dag }a+\frac{\hbar \Omega }{2}\sigma _{0}+\lambda \left(
\sigma _{+}+\sigma _{-}\right) \left( a^{\dag }+a\right) ,  \tag{7}
\end{equation}%
where the parameter $\omega $ is the boson field frequency, $\hbar \Omega $
is the atomic level separation and $\lambda $ is the atom--field coupling
constant. Here $a^{\dag }\left( \text{\textit{resp}. }a\right) $ is a
creation (\textit{resp}. annihilation) operator for the field mode and the $%
\sigma $ are the usual Pauli spin matrices $\sigma _{x}=\sigma _{+}+\sigma
_{-}$, $\sigma _{y}=-i\left( \sigma _{+}-\sigma _{-}\right) $ and $\sigma
_{z}=\sigma _{0}$. Note that the zero energy level was taken halfway between
the two atomic levels, so that the unperturbed atomic energies are $\pm 
\frac{\hbar \Omega }{2}$. The four terms appearing in the interaction part
of the Hamiltonian (7) have the following interpretation: $\sigma _{+}a$ $%
\left( \text{\textit{resp.} }\sigma _{-}a^{\dag }\right) $ represents an
absorbed (\textit{resp.} emitted) photon and an excited (\textit{resp.}
de--excited) atom from state $\left\vert 1\right\rangle $ to $\left\vert
2\right\rangle $, while $\sigma _{+}a^{\dag }$ $\left( \text{\textit{resp.} }%
\sigma _{-}a\right) $ stands for one photon which is emitted (\textit{resp.}
absorbed) and an excited (\textit{resp}. de--excited) atom.

In this present paper, we propose an alternative approach to study the Rabi
Hamiltonian (7) based to the AIM procedure, to obtain both the energy
eigenvalues $\mathcal{E}_{n}$, the eigenfunctions $\chi \left( \xi ;\mathcal{%
E}\right) $ and a set of orthogonal polynomials $\mathcal{P}_{n}\left( 
\mathcal{E}\right) $ explicitly constructed from the above procedure and
known to be orthogonal polynomials of Bender--Dunne of the genus one. In
particular, we show firstly that these orthogonal polynomials generated from
the three--term recursion relation and the coefficients of the polynomial
solutions deduced starting from AIM procedure are equivalent, and secondly
that the orthogonal polynomials $\mathcal{P}_{n}\left( \mathcal{E}\right) $
are prone with a nonpositive definite norm, unless the constraint $n\geq 
\mathcal{E}+\lambda ^{2}+1$ is checked.

The paper proceeds as the following. In the next section we introduce a
unitary transformation with an aim of making the Rabi Hamiltonian simpler.
By applying AIM procedure, we deduce the coefficients of the polynomial
solutions with the first five iterations. In section 3 we clarify, through
an explicit construction, the relation between the calculated eigenfunctions
and the orthogonal polynomials leading to the three-term recursion relation
and which enables us to deduce the orthogonal polynomials $\mathcal{P}%
_{n}\left( \mathcal{E}\right) $. We show, in section 4, that the orthogonal
polynomials are associated with a nonpositive norm $\gamma _{n}$ and the
last section is devoted to our final conclusion.

\section{Solvability of the Rabi Hamiltonian}

In Fock--Bargmann space, the system (7) is transformed into the third--order
differential equation which is not easy to solve. Therefore, the first step
to solve the system (7) is to reduce the order of the differential equation
using some transformations. The structure of the Rabi Hamiltonian is more
easily seen by applying the unitary transformation:%
\begin{eqnarray}
\mathcal{H} &\mathcal{=}&U^{\dag }HU  \notag \\
&=&a^{\dag }a+\beta \left( \sigma _{+}+\sigma _{-}\right) +\lambda \sigma
_{0}\left( a^{\dag }+a\right) ,  \TCItag{8}
\end{eqnarray}%
where $U\equiv U^{\dag }=\frac{1}{\sqrt{2}}\left( \sigma _{+}+\sigma
_{-}+\sigma _{0}\right) .$ Here, $\hbar $ and $\omega $ are set to unity,
and $\Omega =2\beta $. By performing a transition to a Fock--Bargmann space $%
a^{\dag }\rightarrow x$ and $a\rightarrow \frac{d}{dx}$, and writing the
stationary Schr\"{o}dinger equation for the two--component wave--function $%
\tbinom{\psi _{1}\left( x\right) }{\psi _{2}\left( x\right) }$, we obtain a
system of two first order linear differential equations for the functions $%
\psi _{1}\left( x\right) $ and $\psi _{2}\left( x\right) $ \cite{19,20}:%
\begin{eqnarray}
\frac{d\psi _{1}\left( x\right) }{dx} &=&\frac{\mathcal{E}-\lambda x}{%
x+\lambda }\psi _{1}\left( x\right) -\frac{\beta }{x+\lambda }\psi
_{2}\left( x\right) ,  \TCItag{9.1} \\
\frac{d\psi _{2}\left( x\right) }{dx} &=&-\frac{\beta }{x-\lambda }\psi
_{1}\left( x\right) +\frac{\mathcal{E}+\lambda x}{x-\lambda }\psi _{2}\left(
x\right) ,  \TCItag{9.2}
\end{eqnarray}%
where $\mathcal{E}$ are energy eigenvalues of $\mathcal{H}$. The general
solution of Eqs.(9) can be obtained by transforming them into a
second--order form:%
\begin{multline}
\left( x^{2}-\lambda ^{2}\right) \psi _{1}^{\prime \prime }\left( x\right) - 
\left[ \left( \mathcal{E}-\lambda x-1\right) \left( x-\lambda \right)
+\left( x+\lambda \right) \left( \mathcal{E}+\lambda x\right) \right] \psi
_{1}^{\prime }\left( x\right)  \notag \\
-\left[ \beta ^{2}-\mathcal{E}^{2}+\lambda ^{2}x^{2}-\lambda \left(
x-\lambda \right) \right] \psi _{1}\left( x\right) =0,  \tag{10}
\end{multline}%
and substituting%
\begin{equation}
x=\lambda \left( 2\xi -1\right) ,\quad \psi _{1}\left( x\right)
=e^{-2\lambda ^{2}\xi }\chi \left( \xi \right) ,  \tag{11}
\end{equation}%
we obtain a second--order differential equation \cite{19}:%
\begin{multline}
\xi \left( 1-\xi \right) \chi ^{\prime \prime }\left( \xi \right) +\left[
\lambda ^{2}\left( 4\xi ^{2}-2\xi -1\right) +\mathcal{E}\left( 2\xi
-1\right) -\xi +1\right] \chi ^{\prime }\left( \xi \right)  \notag \\
+\left[ \lambda ^{4}\left( 3-4\xi \right) +2\mathcal{E}\lambda ^{2}\left(
1-2\xi \right) +\beta ^{2}-\mathcal{E}^{2}\right] \chi \left( \xi \right) =0.
\tag{12}
\end{multline}

Now this equation is similar to Eq.(4) and then it is in suitable form for
application of AIM procedure; the initial $r_{0}\left( \xi \right) $ and $%
s_{0}\left( \xi \right) $ functions are given:%
\begin{eqnarray}
r_{0}\left( \xi \right) &=&\frac{\lambda ^{2}\left( 4\xi ^{2}-2\xi -1\right)
+\mathcal{E}\left( 2\xi -1\right) -\xi +1}{\xi \left( \xi -1\right) }, 
\TCItag{13.1} \\
s_{0}\left( \xi \right) &=&\frac{\lambda ^{4}\left( 3-4\xi \right) +2%
\mathcal{E}\lambda ^{2}\left( 1-2\xi \right) +\beta ^{2}-\mathcal{E}^{2}}{%
\xi \left( \xi -1\right) }.  \TCItag{13.2}
\end{eqnarray}

We may calculate $r_{n}\left( \xi \right) $ and $s_{n}\left( \xi \right) $
using a sequence of Eqs.(5) and the calculated energy eigenvalues $\mathcal{E%
}_{n}$ by the mean of Eq.(6) should be independent of the choice of the
variable $\xi $.

We report below the factored form of the first five iterations of Eq.(12)
which lead to the coefficients of the polynomial solutions. By convention $n$
represents the iteration number, while $d$ refers to the degree of the
polynomial solutions. The coefficients $\mathcal{Y}_{n}\left( \mathcal{E}%
;\lambda ,\beta \right) $ are polynomials in the energy variable $\mathcal{E}%
_{n}$ with an even higher--degree $2n+2$.

Note that the energy eigenvalue index is the same as the iteration number.

\begin{center}
\begin{tabular}{ccl}
$n=1,$ & $d=2,$ & $\mathcal{C}_{12}=16\lambda ^{4}\left( -1+\mathcal{E}%
+\lambda ^{2}\right) \left( \mathcal{E}+\lambda ^{2}\right) $ \\ 
& $d=1,$ & $\mathcal{C}_{11}=8\lambda ^{2}\left( -1+\mathcal{E}+\lambda
^{2}\right) \mathcal{Y}_{0}\left( \mathcal{E};\lambda ,\beta \right) $ \\ 
& $d=0,$ & $\mathcal{C}_{10}=\mathcal{Y}_{1}\left( \mathcal{E};\lambda
,\beta \right) $ \\ 
$n=2,$ & $d=3,$ & $\mathcal{C}_{23}=64\lambda ^{6}\left( -2+\mathcal{E}%
+\lambda ^{2}\right) \left( -1+\mathcal{E}+\lambda ^{2}\right) \left( 
\mathcal{E}+\lambda ^{2}\right) $ \\ 
& $d=2,$ & $\mathcal{C}_{22}=48\lambda ^{4}\left( -2+\mathcal{E}+\lambda
^{2}\right) \left( -1+\mathcal{E}+\lambda ^{2}\right) \mathcal{Y}_{0}\left( 
\mathcal{E};\lambda ,\beta \right) $ \\ 
& $d=1,$ & $\mathcal{C}_{21}=12\lambda ^{2}\left( -2+\mathcal{E}+\lambda
^{2}\right) \mathcal{Y}_{1}\left( \mathcal{E};\lambda ,\beta \right) $ \\ 
& $d=0,$ & $\mathcal{C}_{20}=\mathcal{Y}_{2}\left( \mathcal{E};\lambda
,\beta \right) $ \\ 
$n=3,$ & $d=4,$ & $\mathcal{C}_{34}=256\lambda ^{8}\left( -3+\mathcal{E}%
+\lambda ^{2}\right) \left( -2+\mathcal{E}+\lambda ^{2}\right) \left( -1+%
\mathcal{E}+\lambda ^{2}\right) \left( \mathcal{E}+\lambda ^{2}\right) $ \\ 
& $d=3,$ & $\mathcal{C}_{33}=256\lambda ^{6}\left( -3+\mathcal{E}+\lambda
^{2}\right) \left( -2+\mathcal{E}+\lambda ^{2}\right) \left( -1+\mathcal{E}%
+\lambda ^{2}\right) \mathcal{Y}_{0}\left( \mathcal{E};\lambda ,\beta
\right) $ \\ 
& $d=2,$ & $\mathcal{C}_{32}=96\lambda ^{4}\left( -3+\mathcal{E}+\lambda
^{2}\right) \left( -2+\mathcal{E}+\lambda ^{2}\right) \mathcal{Y}_{1}\left( 
\mathcal{E};\lambda ,\beta \right) $ \\ 
& $d=1,$ & $\mathcal{C}_{31}=16\lambda ^{2}\left( -3+\mathcal{E}+\lambda
^{2}\right) \mathcal{Y}_{2}\left( \mathcal{E};\lambda ,\beta \right) $ \\ 
& $d=0,$ & $\mathcal{C}_{30}=\mathcal{Y}_{3}\left( \mathcal{E};\lambda
,\beta \right) $ \\ 
$n=4,$ & $d=5,$ & $\mathcal{C}_{45}=1024\lambda ^{10}\left( -4+\mathcal{E}%
+\lambda ^{2}\right) \left( -3+\mathcal{E}+\lambda ^{2}\right) \left( -2+%
\mathcal{E}+\lambda ^{2}\right) $ \\ 
&  & $\ \ \ \ \ \ \ \ \ \left( -1+\mathcal{E}+\lambda ^{2}\right) \left( 
\mathcal{E}+\lambda ^{2}\right) $ \\ 
& $d=4,$ & $\mathcal{C}_{44}=1280\lambda ^{8}\left( -4+\mathcal{E}+\lambda
^{2}\right) \left( -3+\mathcal{E}+\lambda ^{2}\right) \left( -2+\mathcal{E}%
+\lambda ^{2}\right) $ \\ 
\multicolumn{1}{l}{} & \multicolumn{1}{l}{} & $\ \ \ \ \ \ \ \ \ \left( -1+%
\mathcal{E}+\lambda ^{2}\right) \mathcal{Y}_{0}\left( \mathcal{E};\lambda
,\beta \right) $ \\ 
& $d=3,$ & $\mathcal{C}_{43}=640\lambda ^{6}\left( -4+\mathcal{E}+\lambda
^{2}\right) \left( -3+\mathcal{E}+\lambda ^{2}\right) \left( -2+\mathcal{E}%
+\lambda ^{2}\right) \mathcal{Y}_{1}\left( \mathcal{E};\lambda ,\beta
\right) $ \\ 
& $d=2,$ & $\mathcal{C}_{42}=160\lambda ^{4}\left( -4+\mathcal{E}+\lambda
^{2}\right) \left( -3+\mathcal{E}+\lambda ^{2}\right) \mathcal{Y}_{2}\left( 
\mathcal{E};\lambda ,\beta \right) $ \\ 
& $d=1,$ & $\mathcal{C}_{41}=20\lambda ^{2}\left( -4+\mathcal{E}+\lambda
^{2}\right) \mathcal{Y}_{3}\left( \mathcal{E};\lambda ,\beta \right) $ \\ 
& $d=0,$ & $\mathcal{C}_{40}=\mathcal{Y}_{4}\left( \mathcal{E};\lambda
,\beta \right) $ \\ 
$n=5,$ & $d=6,$ & $\mathcal{C}_{56}=4096\lambda ^{12}\left( -5+\mathcal{E}%
+\lambda ^{2}\right) \left( -4+\mathcal{E}+\lambda ^{2}\right) \left( -3+%
\mathcal{E}+\lambda ^{2}\right) $ \\ 
\multicolumn{1}{l}{} & \multicolumn{1}{l}{} & $\ \ \ \ \ \ \ \ \ \left( -2+%
\mathcal{E}+\lambda ^{2}\right) \left( -1+\mathcal{E}+\lambda ^{2}\right)
\left( \mathcal{E}+\lambda ^{2}\right) $ \\ 
& $d=5,$ & $\mathcal{C}_{55}=1280\lambda ^{10}\left( -5+\mathcal{E}+\lambda
^{2}\right) \left( -4+\mathcal{E}+\lambda ^{2}\right) \left( -3+\mathcal{E}%
+\lambda ^{2}\right) $ \\ 
\multicolumn{1}{l}{} & \multicolumn{1}{l}{} & $\ \ \ \ \ \ \ \ \ \left( -2+%
\mathcal{E}+\lambda ^{2}\right) \left( -1+\mathcal{E}+\lambda ^{2}\right) 
\mathcal{Y}_{0}\left( \mathcal{E};\lambda ,\beta \right) $ \\ 
& $d=4,$ & $\mathcal{C}_{54}=3840\lambda ^{8}\left( -5+\mathcal{E}+\lambda
^{2}\right) \left( -4+\mathcal{E}+\lambda ^{2}\right) \left( -3+\mathcal{E}%
+\lambda ^{2}\right) $ \\ 
&  & $\ \ \ \ \ \ \ \ \ \left( -2+\mathcal{E}+\lambda ^{2}\right) \mathcal{Y}%
_{1}\left( \mathcal{E};\lambda ,\beta \right) $ \\ 
& $d=3,$ & $\mathcal{C}_{53}=1280\lambda ^{6}\left( -5+\mathcal{E}+\lambda
^{2}\right) \left( -4+\mathcal{E}+\lambda ^{2}\right) \left( -3+\mathcal{E}%
+\lambda ^{2}\right) \mathcal{Y}_{2}\left( \mathcal{E};\lambda ,\beta
\right) $ \\ 
& $d=2,$ & $\mathcal{C}_{52}=240\lambda ^{4}\left( -5+\mathcal{E}+\lambda
^{2}\right) \left( -4+\mathcal{E}+\lambda ^{2}\right) \mathcal{Y}_{3}\left( 
\mathcal{E};\lambda ,\beta \right) $ \\ 
& $d=1,$ & $\mathcal{C}_{51}=24\lambda ^{2}\left( -5+\mathcal{E}+\lambda
^{2}\right) \mathcal{Y}_{4}\left( \mathcal{E};\lambda ,\beta \right) $ \\ 
& $d=0,$ & $\mathcal{C}_{50}=\mathcal{Y}_{5}\left( \mathcal{E};\lambda
,\beta \right) $%
\end{tabular}
\end{center}

These polynomials have some interesting properties. For example, a
polynomial set $\mathcal{Y}_{m}\left( \mathcal{E};\lambda ,\beta \right) $
can be generated successively for each a new iteration with the constraint $%
n=d+m$, where $0<m\leq n$. However if $n=d$ (i.e. $m=0$), then the
associated polynomial solution is:%
\begin{equation}
\mathcal{Y}_{0}\left( \mathcal{E};\lambda ,\beta \right) =\mathcal{E}%
^{2}-2\lambda ^{2}\mathcal{E}-\beta ^{2}-3\lambda ^{4}.  \tag{14}
\end{equation}

The factor associated to the coefficient $\mathcal{C}_{n,n+1}$ is $\left(
4\lambda ^{2}\right) ^{n+1}$ and $\mathcal{C}_{n0}\equiv \mathcal{Y}%
_{n}\left( \mathcal{E};\lambda ,\beta \right) $, with $n=1,2,\ldots $.
Furthermore, we will see that the polynomials $\mathcal{Y}_{n}\left( 
\mathcal{E};\lambda ,\beta \right) $ lead to the Bender--Dunne orthogonal
polynomials as reviewed in the next section.

\section{Bender--Dunne orthogonal polynomials}

The search for a power series solution of Eq.(12), $\chi \left( \xi \right)
=\sum\nolimits_{n=0}^{\infty }\chi _{n}\left( \xi \right) $, can be
explicitly generated by means of the series expansion leading to the
three--term recursion relation.

Using the identities:%
\begin{eqnarray}
\chi \left( \xi \right) &=&\xi ^{q}\sum\limits_{n=0}^{\infty }a_{n}\xi ^{n},
\TCItag{15.1} \\
\chi ^{\prime }\left( \xi \right) &=&\xi ^{q}\sum\limits_{n=0}^{\infty
}\left( n+q\right) a_{n}\xi ^{n-1},  \TCItag{15.2} \\
\chi ^{\prime \prime }\left( \xi \right) &=&\xi
^{q}\sum\limits_{n=0}^{\infty }\left( n+q\right) \left( n+q-1\right)
a_{n}\xi ^{n-2},  \TCItag{15.3}
\end{eqnarray}%
where the exponent $q$ and the coefficients $a_{n}$ are still undetermined,
and by substituting them into Eq.(12), we have:%
\begin{multline}
\sum\limits_{n=0}^{\infty }4\lambda ^{2}\left( n+q-\lambda ^{2}-\mathcal{E}%
\right) a_{n}\xi ^{n+q+1}+\sum\limits_{n=0}^{\infty }\left[ 3\lambda ^{4}-%
\mathcal{E}^{2}+\beta ^{2}+2\mathcal{E}\lambda ^{2}\right.  \notag \\
\left. +\left( n+q\right) \left( 2\mathcal{E}-\lambda ^{2}-n-q\right) \right]
a_{n}\xi ^{n+q}+\sum\limits_{n=0}^{\infty }\left( n+q\right) \left(
n+q-\lambda ^{2}\right) a_{n}\xi ^{n+q-1}=0,  \tag{16}
\end{multline}%
where the coefficients of the left--hand side of Eq.(16) must vanish
individually. The lowest power of $\xi $ appearing in Eq.(16) is $\xi
^{q-1}, $ for $n=0$ in the last summation. The requirement that the
coefficients vanish yields the \textit{indicial equation }$q\left( q-\lambda
^{2}\right) \,a_{0}=0$; hence, we must require either $q=0$ or $q=\lambda
^{2}$, and $a_{0}\neq 0$. Note that if the first summation vanishes, the
energy eigenvalues satisfy the constraint:%
\begin{equation}
\mathcal{E}_{n,q}\left( \lambda \right) =n+q-\lambda ^{2}.  \tag{17}
\end{equation}

Therefore, if we replace $n\rightarrow j$ in the second summation and $%
\left( n-1\right) \rightarrow j$ in the last (they are independent
summations), this results in the two--term recursion relation:%
\begin{equation}
a_{j+1}=-\frac{\beta ^{2}}{\left( j+q+1\right) \left( j+q+1-\lambda
^{2}\right) }a_{j}.  \tag{18}
\end{equation}

Substituting $j$ by $n$ and choosing the indicial equation root $q=0$, the
mathematical induction leads us to write:%
\begin{equation}
a_{n}=\frac{\left( -1\right) ^{n}\beta ^{2n}}{n!}\frac{\Gamma \left(
1-\lambda ^{2}\right) }{\Gamma \left( n+1-\lambda ^{2}\right) }a_{0}, 
\tag{19}
\end{equation}%
with $a_{0}=1$. The solution of Eq.(12), following Eq.(1)%
\begin{eqnarray}
\chi \left( \xi ;\mathcal{E}\right) &=&\sum\limits_{n=0}^{\infty }\chi
_{n}\left( \xi \right) \mathcal{P}_{n}\left( \mathcal{E}\right)  \notag \\
&=&\sum\limits_{n=0}^{\infty }\frac{\left( -1\right) ^{n}\beta ^{2n}}{n!}%
\frac{\Gamma \left( 1-\lambda ^{2}\right) }{\Gamma \left( n+1-\lambda
^{2}\right) }\mathcal{P}_{n}\left( \mathcal{E}\right) \xi ^{n},  \TCItag{20}
\end{eqnarray}%
can be considered as the generating function for the polynomials $\mathcal{P}%
_{n}\left( \mathcal{E}\right) $. However, for positive integer values $%
\Lambda $, the series expansion in Eq.(20) are truncated when $\mathcal{E}$
is a zero of $\mathcal{P}_{\Lambda }\left( \mathcal{E}\right) $. We know
that $\Gamma \left( n+1-\lambda ^{2}\right) $ has simple poles at $n=\lambda
^{2}-1,\lambda ^{2}-2,\lambda ^{2}-3,\ldots $, \cite{24} and taking into
account the constraint (17) with $q=0$, leading to the upper limit for $n$,
i.e. $n_{\max }\equiv \Lambda =\mathcal{E}+\lambda ^{2}>0$. Indeed, this can
be observed in the coefficients $\mathcal{C}_{nd}$\ of the polynomial
solutions like it was repoted above. Therefore, the corresponding exact
eigenfunctions are given:%
\begin{equation}
\chi \left( \xi ;\mathcal{E}\right) =\sum\limits_{n=0}^{\Lambda }\frac{%
\left( -1\right) ^{n}\beta ^{2n}}{n!}\frac{\Gamma \left( 1-\lambda
^{2}\right) }{\Gamma \left( n+1-\lambda ^{2}\right) }\mathcal{P}_{n}\left( 
\mathcal{E}\right) \xi ^{n}.  \tag{21}
\end{equation}

By substituting Eq.(21) into Eq.(12) leads to the following three--term
recursion relation for $\mathcal{P}_{n}\left( \mathcal{E}\right) $:%
\begin{multline}
4\lambda ^{2}\left( n-\mathcal{E}-\lambda ^{2}\right) \mathcal{P}%
_{n+1}\left( \mathcal{E}\right) +\left[ \left( n-\mathcal{E}-\lambda
^{2}\right) \left( \mathcal{E}-3\lambda ^{2}-n\right) -\beta ^{2}\right] 
\mathcal{P}_{n}\left( \mathcal{E}\right)  \notag \\
+4\lambda ^{2}\left( n-\mathcal{E}-\lambda ^{2}\right) \mathcal{P}%
_{n-1}\left( \mathcal{E}\right) =0,  \tag{22}
\end{multline}%
with the normalization condition $\mathcal{P}_{0}\left( \mathcal{E}\right)
=1 $. The polynomial $\mathcal{P}_{n}\left( \mathcal{E}\right) $ vanishes
for $n\geq \mathcal{E}+\lambda ^{2}+1$ as occurred in the third term of
Eq.(22). The recursion relation (22) generates a set of polynomials, where
the first five of them are reported below in table 1.

\begin{table}[h] \centering%
\begin{tabular}{c||c||l}
\hline\hline
$n$ & $\mathcal{E}\left( \lambda \right) $ & $\mathcal{P}_{n}\left( \lambda
\right) \QQfnmark{%
Actually, the deduced orthogonal polynomials are written as a product of $%
\mathcal{P}_{n}\left( \lambda \right) $ by $\left( -1\right) ^{n-1}\beta
^{2} $.}$ \\ \hline
$1$ & $1-\lambda ^{2}$ & $4\lambda ^{2}+\left( \beta ^{2}-1\right) $ \\ 
\hline
$2$ & $2-\lambda ^{2}$ & $32\lambda ^{4}+4\left( 3\beta ^{2}-8\right)
\lambda ^{2}+\left( \beta ^{2}-1\right) \left( \beta ^{2}-4\right) $ \\ 
\hline
$3$ & $3-\lambda ^{2}$ & $384\lambda ^{6}+16\left( 11\beta ^{2}-54\right)
\lambda ^{4}+8\left( 3\beta ^{4}-29\beta ^{2}+54\right) \lambda ^{2}$ \\ 
&  & $+\left( \beta ^{2}-1\right) \left( \beta ^{2}-4\right) \left( \beta
^{2}-9\right) $ \\ \hline
$4$ & $4-\lambda ^{2}$ & $6144\lambda ^{8}+128\left( 25\beta ^{2}-192\right)
\lambda ^{6}+16\left( 35\beta ^{4}-542\beta ^{2}+1728\right) \lambda
^{4}+8\left( 5\beta ^{6}\right. $ \\ 
&  & $\left. -115\beta ^{4}+722\beta ^{2}-1152\right) \lambda ^{2}+\left(
\beta ^{2}-1\right) \left( \beta ^{2}-4\right) \left( \beta ^{2}-9\right)
\left( \beta ^{2}-16\right) $ \\ \hline
$5$ & $5-\lambda ^{2}$ & $122880\lambda ^{10}+512\left( 137\beta
^{2}-1500\right) \lambda ^{8}+64\left( 225\beta ^{4}-5036\beta
^{2}+24000\right) \lambda ^{6}$ \\ 
&  & $+16\left( 85\beta ^{6}-2867\beta ^{4}+27518\beta ^{2}-72000\right)
\lambda ^{4}+4\left( 15\beta ^{8}-670\beta ^{6}+9551\beta ^{4}\right. $ \\ 
&  & $\left. -49216\beta ^{2}+72000\right) \lambda ^{2}+\left( \beta
^{2}-1\right) \left( \beta ^{2}-4\right) \left( \beta ^{2}-9\right) \left(
\beta ^{2}-16\right) \left( \beta ^{2}-25\right) $ \\ \hline\hline
\end{tabular}%
\QQfntext{0}{
Actually, the deduced orthogonal polynomials are written as a product of $%
\mathcal{P}_{n}\left( \lambda \right) $ by $\left( -1\right) ^{n-1}\beta
^{2} $.}%
\caption{Bender--Dunne orthogonal polynomials for Rabi Hamiltonian and
the associated energy eigenvalues}\label{TableKey}%
\end{table}%

These orthogonal polynomials are exactly similar to those obtained by the
Juddian isolated method and agree with the first roots of the Ku\'{s} series 
\cite{18,19}. On the other word, the system has energy $\mathcal{E}%
_{n}=n-\lambda ^{2},$ $n=1,2,\ldots ,$ only if the atomic--level separation $%
2\beta $ and the boson--atom field coupling $\lambda $ obey condition $%
\mathcal{P}_{n}\left( \lambda \right) =0$. Note also that these orthogonal
polynomials can be deduced starting from the coefficients $\mathcal{C}_{nd}$
of the polynomial solutions $\mathcal{Y}_{n}\left( \mathcal{E};\lambda
,\beta \right) $ with the same associated energy eigenvalues (17).

\section{Norms of orthogonal polynomials}

As the polynomials $\mathcal{P}_{n}\left( \mathcal{E}\right) $ are
orthogonal of a discrete variable $\mathcal{E}_{n},$ it is possible to
determine their squared norms. The procedure consists to apply Eq.(3); i.e.
multiplying the recursion relation (22) by $\mathcal{E}^{n-1}\rho \left( 
\mathcal{E}\right) $, where $\rho \left( \mathcal{E}\right) $ is the weight
function, and to integrate with respect to $\mathcal{E}$ using the fact that
both $\mathcal{P}_{n}\left( \mathcal{E}\right) $ and $\mathcal{E}^{k}$, $k<n$%
, are orthogonal. We obtain:%
\begin{equation}
\gamma _{n}=2\lambda ^{2}\frac{n-\lambda ^{2}}{n+\lambda ^{2}}\gamma _{n-1},
\tag{23}
\end{equation}%
with $\gamma _{n}=p_{n}^{2}$ and $\gamma _{0}=1$. It is obvious that by
mathematical induction, we get:%
\begin{eqnarray}
\gamma _{n} &=&\prod\limits_{k=1}^{n}2\lambda ^{2}\frac{k-\lambda ^{2}}{%
k+\lambda ^{2}}\gamma _{0}  \notag \\
&=&2^{n}\lambda ^{2n}\frac{\left( 1-\lambda ^{2}\right) _{n}}{\left(
1+\lambda ^{2}\right) _{n}},  \TCItag{24}
\end{eqnarray}%
where $\left( a\right) _{n}=a\left( a+1\right) \left( a+2\right) \cdots
\left( a+n-1\right) $ are the often--used Pochhammer symbol \cite{24}, with $%
\left( a\right) _{0}=1$. Using the well--known identity $\Gamma \left(
z\right) \Gamma \left( 1-z\right) =\frac{\pi }{\sin \pi z}$ \cite{24}, we
have:%
\begin{equation}
\gamma _{n}=2^{n}\lambda ^{2n}\,\frac{\sin \pi \lambda ^{2}}{\pi }\Gamma
\left( \lambda ^{2}\right) \Gamma \left( 1+\lambda ^{2}\right) \frac{\Gamma
\left( n+1-\lambda ^{2}\right) }{\Gamma \left( n+1+\lambda ^{2}\right) }. 
\tag{25}
\end{equation}

This identity reveals an important result that the orthogonal polynomials
are associated with a nonpositive definite norm represented by $\Gamma
\left( n+1+\lambda ^{2}\right) $. However, it shows in the $\Gamma $%
--function of the numerator that $\gamma _{n}$ has in the complex plan poles
which are given at the negative integers $N=0,-1,-2,\ldots $, (or
equivalently: $n=\lambda ^{2}-1,\lambda ^{2}-2,\ldots $), leading to the
upper limit $\Lambda =\mathcal{E}+\lambda ^{2}>0$ as conjectured above.
Consequently, $\gamma _{n}$ vanishes for each $n\geq \Lambda +1$.

\section{Conclusion}

In this paper, we have shown that there exists another approach of
introducing the Bender--Dunne polynomials which characterizes the
quasi-exact solvable problems. The model given here is related to the AIM
procedure and we have taken a new look at the solution of the Rabi
Hamiltonian in order to generate its energy eigenvalues as well as the
eigenfunctions.

The main aim of this article was to consider the possibility of generating
the Bender--Dunne polynomials $\mathcal{P}_{n}\left( \mathcal{E}\right) $\
via two different but equivalent ways. The first way, considered as the
standard method, consists in applying the eigenfunctions to the three--term
recursion relation, while the second enables us to generate them directly
starting from the coefficients $\mathcal{C}_{nd}$ of the polynomial
solutions $\mathcal{Y}_{n}\left( \mathcal{E};\lambda ,\beta \right) $
deduced by AIM procedure. Consequently, it proves that $\mathcal{P}%
_{n}\left( \mathcal{E}\right) $ are similar to $\mathcal{Y}_{n}\left( 
\mathcal{E};\lambda ,\beta \right) $\ if the system has the associated
energy eigenvalues $\mathcal{E}_{n}=n-\lambda ^{2}$, $n=1,2,\ldots $ We have
also shown that the orthogonal polynomials arising from a quasi--exactly
solvable Rabi model is associated with a nonpositive definite norm whose the
iteration number $n$ exceeds a critical value $\Lambda $ mentioned above,
i.e. $n\geq \Lambda +1$. Consequently, the quasi--exact energy eigenvalues
of the Rabi Hamiltonian are the zeros of the polynomials $\mathcal{P}%
_{\Lambda }\left( \mathcal{E}\right) $.

For other applications, we believe that similar results can be obtained via
the approach worked out here to the various atomic systems such as
Jaynes--Cummings and $\mathcal{E}\otimes \epsilon $ Jahn--Teller
Hamiltonians.

\end{document}